\documentclass[twocolumn,prl,aps,superscriptaddress,showpacs]{revtex4-1}

\usepackage{float}

\usepackage{mathptmx}
\usepackage{dcolumn}
\usepackage{amsmath}
\usepackage{dsfont}
\usepackage{bm}
\usepackage{color}
\usepackage{latexsym}
\usepackage{epstopdf}
\usepackage[english]{babel}
\usepackage{enumerate}

\usepackage[pdftex]{graphicx}

\usepackage{natbib}
\usepackage{epstopdf}
\usepackage{appendix}

\definecolor{mygrey}{gray}{0.45}
\definecolor{myblue}{rgb}{0.2,0.2,0.8}
\definecolor{myzard}{cmyk}{0,0,0.05,0}
\definecolor{mywhite}{rgb}{1,1,1}
\definecolor{mywhite}{rgb}{1,1,1}
\definecolor{myred}{rgb}{1,0.,0.3}
\usepackage[colorlinks=true,citecolor=myblue,linkcolor=myred,hidelinks=true]{hyperref}

\newcommand{\bra}[1]{\left\langle #1\right|}
\newcommand{\ket}[1]{\left| #1\right\rangle}

\newcommand\kk{\mathbf{k}}
\newcommand\rr{\mathbf{r}}

\begin{document}

\title{Tunable and robust long-range coherent interactions between quantum emitters mediated by Weyl bound states}
\author{I\~naki Garc\'{i}a-Elcano}
\affiliation{Departamento de F\'{i}sica Te\'{o}rica de la Materia Condensada  and Condensed Matter Physics Center (IFIMAC), Universidad Aut\'{o}noma de Madrid, E-28049 Madrid, Spain}
\author{Alejandro Gonz\'alez-Tudela}
\email{a.gonzalez.tudela@csic.es}
\affiliation{Instituto de F\'{i}sica Fundamental IFF-CSIC, Calle Serrano 113b, Madrid 28006, Spain}
\author{Jorge Bravo-Abad}
\email{jorge.bravo@uam.es}
\affiliation{Departamento de F\'{i}sica Te\'{o}rica de la Materia Condensada  and Condensed Matter Physics Center (IFIMAC), Universidad Aut\'{o}noma de Madrid, E-28049 Madrid, Spain}

\begin{abstract}
	
Long-range coherent interactions between quantum emitters are instrumental for quantum information and simulation technologies, but they are generally difficult to isolate from dissipation. Here, we show how such interactions can be obtained in photonic Weyl environments due to the emergence of an exotic bound state whose wavefunction displays power-law spatial confinement. Using an exact formalism, we show how this bound state can mediate coherent transfer of excitations between emitters, with virtually no dissipation and with a transfer rate that follows the same scaling with distance as the bound state wavefunction. In addition, we show that the topological nature of Weyl points translates into two important features of the Weyl bound state, and consequently of the interactions it mediates: first, its range can be tuned without losing the power-law confinement, and, second, they are robust under energy disorder of the bath. To our knowledge, this is the first proposal of a photonic setup that combines simultaneously coherence, tunability, long-range, and robustness to disorder. These findings could ultimately pave the way for the design of more robust long-distance entanglement protocols or quantum simulation implementations for studying long-range interacting systems.
\end{abstract}

\maketitle

Obtaining long-distance coherent interactions is one of the current frontiers in atomic physics. Such interactions can be harnessed, for example, to induce long-distance  entanglement between emitters~\cite{shahmoon13a}, create large optical non-linearities~\cite{shahmoon16a}, or study long-range interacting many-body physics in the context of quantum simulation~\cite{hauke13a,richerme14a,maghrebi16a,koffel12a,vodola14a,kastner11a,blass18a,arguello19a}, where they are known to lead to qualitatively different physics than short-ranged interactions. Unfortunately, these long-range coherent interactions are difficult to isolate from dissipation. For example, free-space photons can mediate coherent interactions between emitters scaling with the distance ($r$) as $\propto1/r^3$. However, these interactions are unavoidably accompanied by individual and collective dissipation that scales in a similar fashion~\cite{lehmberg70a,lehmberg70b}. An elegant way of overcoming this fundamental problem consists in modifying the photonic environment around the emitters~\cite{purcell46a}, e.g., by placing the emitters in gapped electromagnetic media like photonic-crystals~\cite{bykov75a,john90a,kurizki90a}. When the emitter's transition frequency lies within a photonic band-gap, the vanishing density states cancels dissipation, while localizing the photonic excitations around the emitter, forming a so-called atom-photon bound-state (BS)~\cite{cohenbook92a}. Remarkably, these BSs can still mediate coherent interactions through the overlap of the BS wavefunction of the different emitters.~Although tunable through the detuning  between the emitter's transition frequency and the band-edge~\cite{douglas15a,gonzaleztudela15c,hauke13a,richerme14a,maghrebi16a,koffel12a,vodola14a,kastner11a,blass18a,arguello19a}, photonic band-gap BSs are exponentially localized, losing the power-law scaling of the interactions.~In systems featuring two-dimensional photonic Dirac points, one can recover the power-law scaling for the interaction~\cite{gonzaleztudela18c,perczel18a}, but at the expense of sacrificing the tunability of the interactions.

In this Letter, we show that photonic Weyl environments~\cite{lu15a,bravo15a,dubcek15a,xiao16b, lin16a, chen16a, gao16a, noh17a,yang18b,roy18a,yang18a,jia19a} can mediate coherent interactions between quantum emitters featuring negliglible dissipation, power-law scaling, tunability, and robustness to disorder. By using a fully non-perturbative approach, we find that, when the frequency of the emitter matches that of the Weyl point, an exotic BS with power-law localization emerges around the emitter. This finding is in stark contrast to a recent study reporting that the strength of light-matter interaction vanishes when the emitter's transition is tuned to the Weyl point frequency~\cite{ying19a}. In addition, by computing the exact dynamics of two emitters, we numerically corroborate that this Weyl bound state can mediate coherent power-law interactions between the emitters with virtually no dissipation. Finally, we show that this bound state wavefunction, and the corresponding interaction between QEs it mediates, inherit two important features from the topological protection of Weyl points~\cite{armitage18a}. First, the photonic band-structure around the Weyl points can be modified without opening a band-gap; this enables tuning the power-law exponent of the interaction $1/r^\alpha$, with $\alpha$ taking values in the interval $[3/2,3]$, depending on the configuration of the system. Second, this power-law behaviour is robust to certain degree of disorder in the bath, as it occurs in other topological BSs~\cite{bello19a,kim20,leonforte20}. The combination of all these features in the same platform (i.e., a single platform enabling coherent, power-law, no dissipative, tunable and robust to disorder interactions between QEs) has to our knowledge never been predicted or reported in any other photonic environment.

\begin{figure}[t]
   \centering
    \includegraphics[width=\columnwidth]{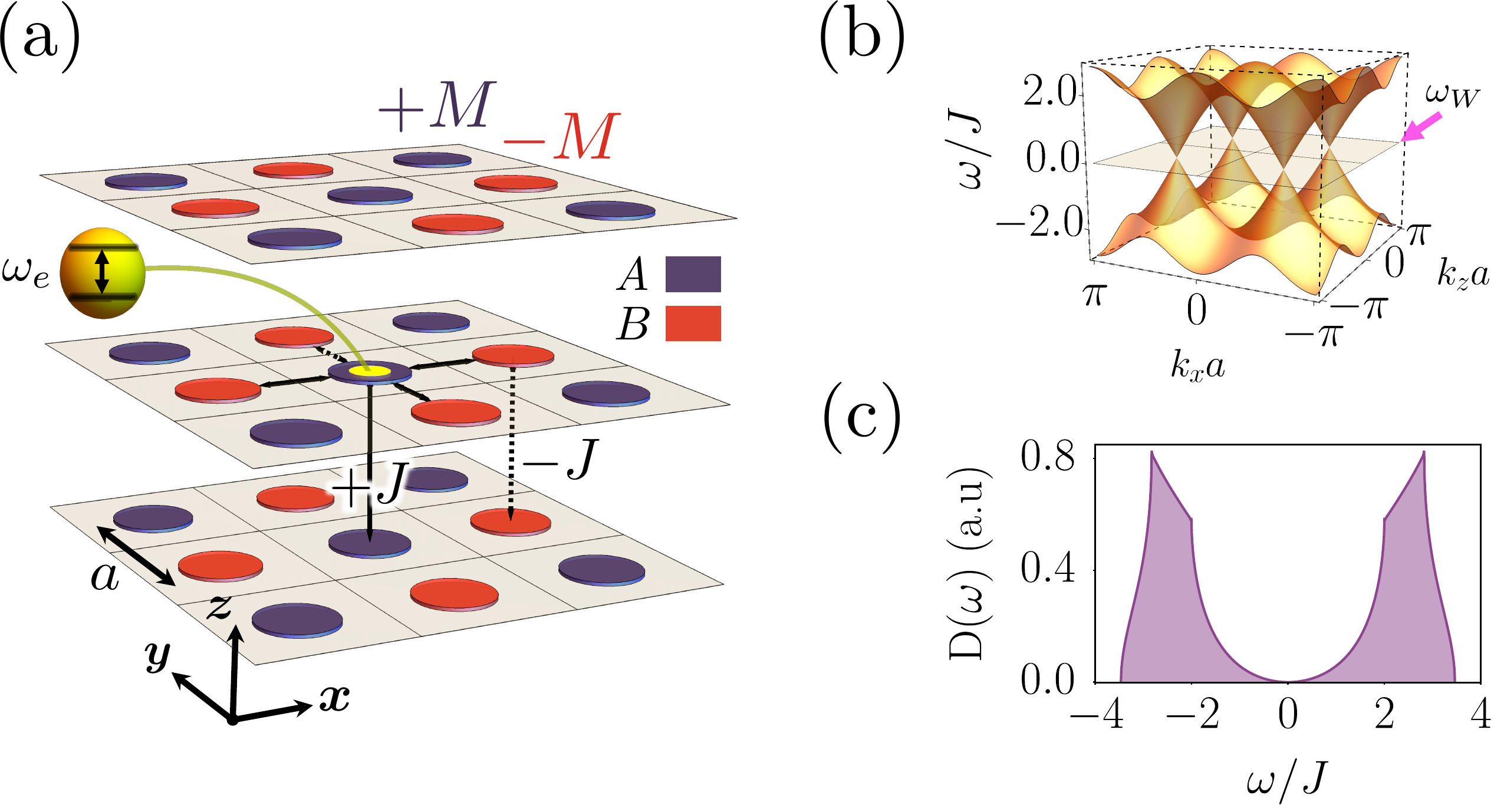}
    \vspace{-0.4cm}
       \caption{
       (a) Schematics of the analyzed system. Blue (red) shallow cylinders represent sites belonging to the $A$ ($B$) sublattice. Solid and dashed arrows correspond to positive and negative first-neighbour hoppings, $J$, respectively. $M$ denotes the sublattice frequency offset, whereas $a$ is the lattice constant. 
       (b) Band structure of the system shown in (a), as calculated for $M=0$. The frequency of the Weyl point ($\omega_W$) is marked with an arrow.
       (c) Density of states, $\mathrm{D}(\omega)$, associated to the bath with $M=0$.
    \vspace{-0.3cm}}
  \label{fig:Fig1}
\end{figure}
 
To illustrate our findings we consider a three-dimensional photonic lattice with the nearest-neighbour hopping pattern of strength $J$ and alternating phases sketched in Fig.~\ref{fig:Fig1}(a). The corresponding bath Hamiltonian can be written as follows (taking $\hbar=1$ and using a rotating frame such that the Weyl point frequency is the energy reference):
\begin{eqnarray}
H_B&=J\,&\sum_{\rr}[(-1)^{x+y}(c_\rr^\dagger c_{\rr+a\hat{z}}-c_\rr^\dagger c_{\rr+a\hat{y}})\nonumber\\
&&+c_\rr^\dagger c_{\rr+a\hat{x}} + \mathrm{H.c.}]+M \sum_\rr(-1)^{x+y}c_\rr^\dagger c_\rr\,,\label{eq:HB}
\end{eqnarray}
where $c_{\rr}^\dagger$ ($c_{\rr}$) creates (annihilates) a bosonic mode at position $\rr$, $a$ is the lattice constant, and $M$ is an alternating on-site energy off-set. The motivation for choosing this discrete lattice model is two-fold. First, it captures well most of the features of the continuum models, while opening the possibility of analytical and numerical understanding of the studied phenomena. Second, this class of discrete lattice models can be readily implemented in recent quantum optical setups based on cold atoms~\cite{devega08a,krinner18a} or coupled microwave resonators~\cite{liu17a,mirhosseini18a}. In this discrete model the photonic timescale is given by $J^{-1}$, while the relevant emitter's interaction timescale scales as $J/g^2$. Thus, the separation of timescales between the two is guaranteed as long as $g< J$. Effects associated to the spatial dependence of the electric field density of the system's eigenmodes or the role of polarization cannot be accounted for by a discrete model and will be addressed in future work.

By imposing periodic boundary conditions, $H_B$ can be diagonalized, leading to two energy bands $\omega_{\pm}(\kk)$ associated to the bipartite nature of the lattice~\cite{SIWeyl} \nocite{bena2009b, guttmann10a, altland2010, abramowitz66a, breuerbook02a, berry1984}. Figures~\ref{fig:Fig1}(b) and (c) display $\omega_{\pm}(\kk)$ and its corresponding density of states $\text{D}(\omega)$, respectively, for $M=0$. Our calculations for $|M|>0$ show that the band structure hosts the expected Weyl points as long as $|M|\leq 2J$,  leading to singular band-gaps at the Weyl frequency. When $|M| = 2J$, Weyl points with opposite chirality meet in the reciprocal space and annihilate in pairs, opening a band-gap around the Weyl frequency that enlarges its width as $|M|$ is further increased.

\begin{figure*}[t]
   \centering
    \includegraphics[width=17cm]{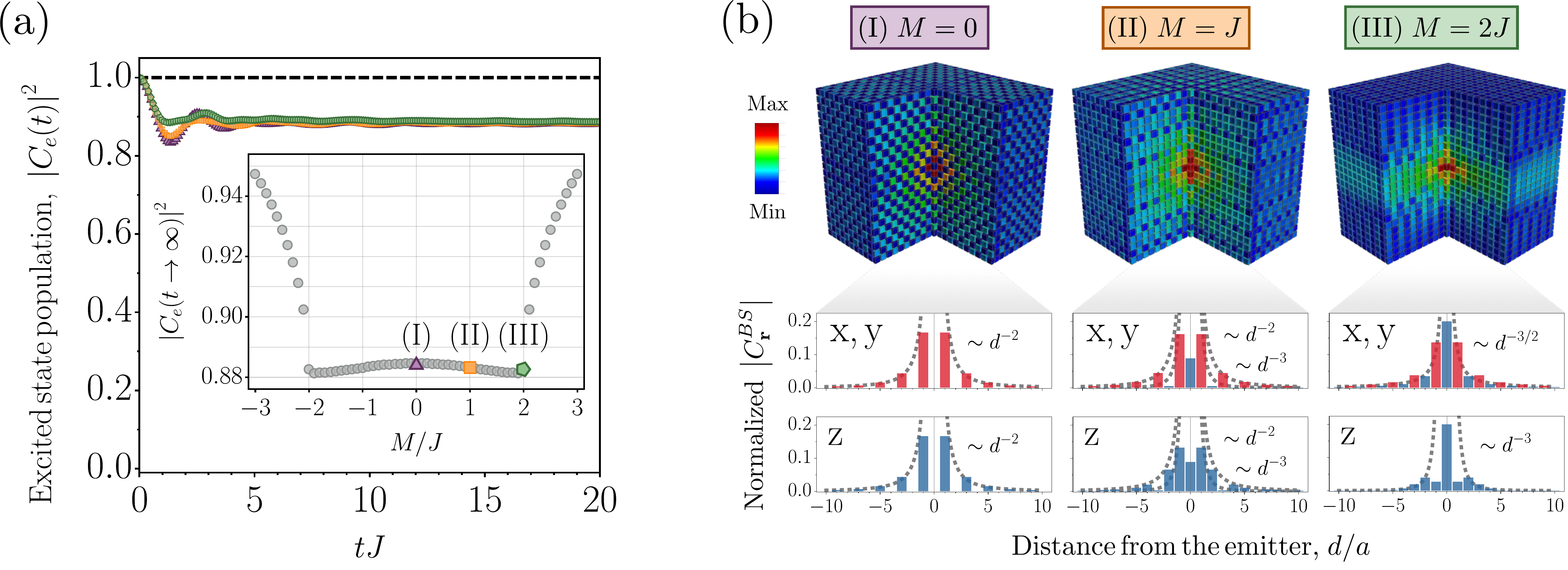}
       \caption{
       (a) Exact dynamics of the emitter's excited state population, $|C_e(t)|^2$, calculated for $M=0$, $M=J$, and $M=2J$ (shown as purple, orange and green lines, respectively). The dashed line stands for the Markovian prediction. Inset shows $|C_e(t\to\infty)|^2$ versus $M$ (the cases displayed in the main figure are marked with the corresponding color code). In all displayed results, $\Delta=\Delta_c$ and $g=0.5 J$ is assumed. (b) Left, center and right upper panels display the three-dimensional distribution of the normalized photonic component of the Weyl bound state for $M=0$, $J$ and $2J$, respectively. Lower panels show cuts of the corresponding spatial distributions along the $x,y$-directions and $z$-direction, as a function of the distance $d$ to the quantum emitter. In these panels, blue (red) bars identify the sites belonging to the $A$ ($B$) sublattice, respectively. For each value of $M$, the emitter is detuned from the Weyl frequency by the associated critical value, $\Delta_c$.
       \vspace{-0.3cm}}
 \label{fig:Fig2}
 \end{figure*}

In this work we are particularly interested in the quantum optical behaviour of quantum emitters interacting through such photonic Weyl environment~\cite{footnote44}. To model each emitter we use a two-level system ($\left\{ \ket{g}, \ket{e} \right\}$), whose transition frequency is assumed to be locally coupled to a specific site of the lattice. Assuming that the light-matter coupling is given by the scalar and real parameter, $g$, the full Hamiltonian of the system can be written as:
\begin{align}
H=H_B+\sum_j \: \Delta \: \sigma^{j}_{ee}+g \left( c_{\rr}\:\sigma^{j}_{eg} +\mathrm{H.c.} \right)\,,\label{eq:HFull}
\end{align}
where $\sigma^{j}_{\mu\nu}=\ket{\mu_j}\bra{\nu_j}$ denotes the \emph{spin operator} of the $j$-th emitter, and $\Delta$ is the detuning between the emitters' transition frequency and the Weyl point. For definiteness, in this Letter we focus on the regime where $|\Delta|\ll J$, such that the emitter only probes the bath's density of states around Weyl frequency. We leave the study of the rest of regions of $\mathrm{D}(\omega)$ (e.g., of the kinks observed in Fig.~1(c)) to future works, since they can also be the source of non-trivial quantum optics phenomena~\cite{gonzaleztudela18d}.

We first analyze the configuration in which a single emitter is coupled to a localized mode of the $A$ sublattice (similar results are obtained when it couples to the $B$ sublattice). We follow the canonical approach used in the literature to study systems hosting BSs, which consists in studying first the dynamical consequences of the BS in spontaneous emission~\cite{john94a,tong10a,longo10a,garmon13a,redchenko14a,lombardo14a,sanchezburillo17a}, and then characterizing its properties using the secular equation of the full Hamiltonian $H$ of Eq.~(\ref{eq:HFull})~\cite{calajo16a,shi16a}. 

The spontaneous emission problem considers an excited emitter with no photons in the bath as the initial state, i.e., $\ket{\Psi_0}=\ket{e}\otimes\ket{\mathrm{vac}}_B$, and then studies its relaxation into the bath due to the interaction with it. The excited state probability amplitude is given by:
\begin{equation}
C_e(t)=\bra{\Psi_0}e^{-i H t}\ket{\Psi_0}\,,
\end{equation}
where $|C_e(t)|^2$ represents the excited state population of the emitter. Perturbative approaches, like Fermi's Golden rule~\cite{fermi32a} or the Born-Markov approximation~\cite{cohenbook92a}, predict just an exponential decay $|C_e(t)|^2\approx e^{-\Gamma t}$, with a decay rate proportional to the density of states, $\Gamma \propto \mathrm{D}(\omega)$. Thus, for an emitter in resonance with the Weyl point of our system ($\Delta=0$), those approaches predict $|C_e(t)|^2=1$, irrespective of the value of $M$ and $J$ (see dashed black line in Fig.~\ref{fig:Fig2}(a)). For the studied problem, however, an exact calculation of the dynamics of $|C_e(t)|^2$ can be done using resolvent operator techniques~\cite{cohenbook92a,nakazato96a} and exploiting the fact that $H$ preserves the number of excitations in the system~\cite{SIWeyl}. 

Main panel of Fig.~\ref{fig:Fig2}(a) shows the calculated exact dynamics of $|C_e(t)|^2$ for a fixed value of $g$ ($g=0.5J$) and three representative values of $M$ ($M=$0, $J$, $2J$). As seen, the results display a rather different behaviour from those expected from a perturbative prediction.~In particular, irrespective of $M$, the emitter initially relaxes until a certain time when the evolution quenches to a finite value $|C_e(t\rightarrow\infty)|^2< 1$. We note that in these calculations, while for $M=0$ the quantum emitter is fully tuned to the Weyl point frequency ($\Delta=0$), for $M\neq 0$ the emitter is slightly detuned from the Weyl point (by $\Delta=\Delta_c$) to compensate for the spectral shift coming from the interaction with the bath~\cite{SIWeyl}. The \emph{fractional decay} of the initial excitation observed in main panel of Fig.~\ref{fig:Fig2}(a) is known to be the dynamical signature of the emergence of a BS in the system~\cite{john94a}, as recently confirmed experimentally in standard photonic band-gap environments~\cite{krinner18a}. 

Remarkably, for the case $M=0$, we found the following analytical expression for the steady-state value of the emitter's population $|C_e(t\rightarrow\infty)|^2\approx \left[ 1+0.25 (g/J)^2 \right]^{-2}$, which unveils the scaling of the non-Markovian correction with the light-matter coupling strength $g$~\cite{SIWeyl} (the perturbative prediction $|C_e(t)|^2 \approx 1$ is recovered for $g/J \ll 1$). On the other hand, inset of Fig.~\ref{fig:Fig2}(a) shows the numerical results for the dependence of the emitter's stationary-state population with $M$ (computed also for $g=0.5J$). As observed,  $|C_e(t\rightarrow\infty)|^2$ is approximately constant for $|M|<2J$ ---the abrupt change observed at $|M| \approx 2J$ corresponds to the band-gap opening in the system. The above findings show the limited value of the results reported in~\cite{ying19a}, which by applying a purely perturbative approach, completely neglect all non-perturbative phenomena associated to light-matter interaction in Weyl-point photonic environments.

Having identified the emergence of a BS in the emitter dynamics, now we turn to characterize its wavefunction (hereafter we refer to this bound state as Weyl Bound State, WBS). Since the WBS exists within the single excitation subspace, its wavefunction can be written as $\ket{\psi_{\mathrm{WBS}}}=(C_e^{\mathrm{WBS}}\,\sigma_{eg}+\sum_{\rr} C_{\rr}^{\mathrm{WBS}}c_\rr^\dagger)\ket{g}\otimes \ket{\mathrm{vac}}_B$, where the probability amplitudes $\left\{C_e^{\mathrm{WBS}},  C_{\rr}^{\mathrm{WBS}} \right\}$ can be obtained by solving the secular equation $H \ket{\psi_{\mathrm{WBS}}} = E_{\mathrm{WBS}} \ket{\psi_{\mathrm{WBS}}}$ for $E_{\mathrm{WBS}}=0$. In the following, we focus on analyzing the projection of the wavefunction over the photonic degrees of freedom, $|C_{\rr}^{WBS}|$. 

Figure~\ref{fig:Fig2}(b) summarizes the obtained spatial distribution for $|C_{\rr}^{WBS}|$, as calculated for $M=0$, $J$ and $2J$. Upper panels of Fig.~\ref{fig:Fig2}(b) render the corresponding three-dimensional visualizations, whereas the lower panels display specific cuts along the $x$, $y$ and $z$ directions. As seen, for $M=0$, the WBS's photonic component localizes around the QE mostly in an isotropic fashion, featuring a spatial decay that fits an inverse square power-law. The physical origin of this power-law confinement, also obtained at longer distances (see Ref.~\cite{SIWeyl}), can be associated to the particular form of the dispersion relation around the Weyl frequency. Moreover, the topological protection of Weyl points implies that, by varying $|M|\in [0,2J)$, one can change locally the band structure without opening a band-gap. This fact brings in the possibility to modify the spatial distribution of the WBS while preserving its power-law character. Middle and right panels of Fig.~\ref{fig:Fig2}(b) show that the power-law exponents governing the bound state's confinement, $1/d^\alpha$, can actually be varied in the interval $\alpha\in [3/2,3]$, depending on $M$ and the spatial direction under consideration. The obtained power-law and tunable spatial decay of the WBS represents a whole new instance of light-matter bound state that, as we discuss below, enables a novel platform for long-distance interaction between quantum emitters.

\begin{figure}[t]
   \centering
    \includegraphics[width=7cm]{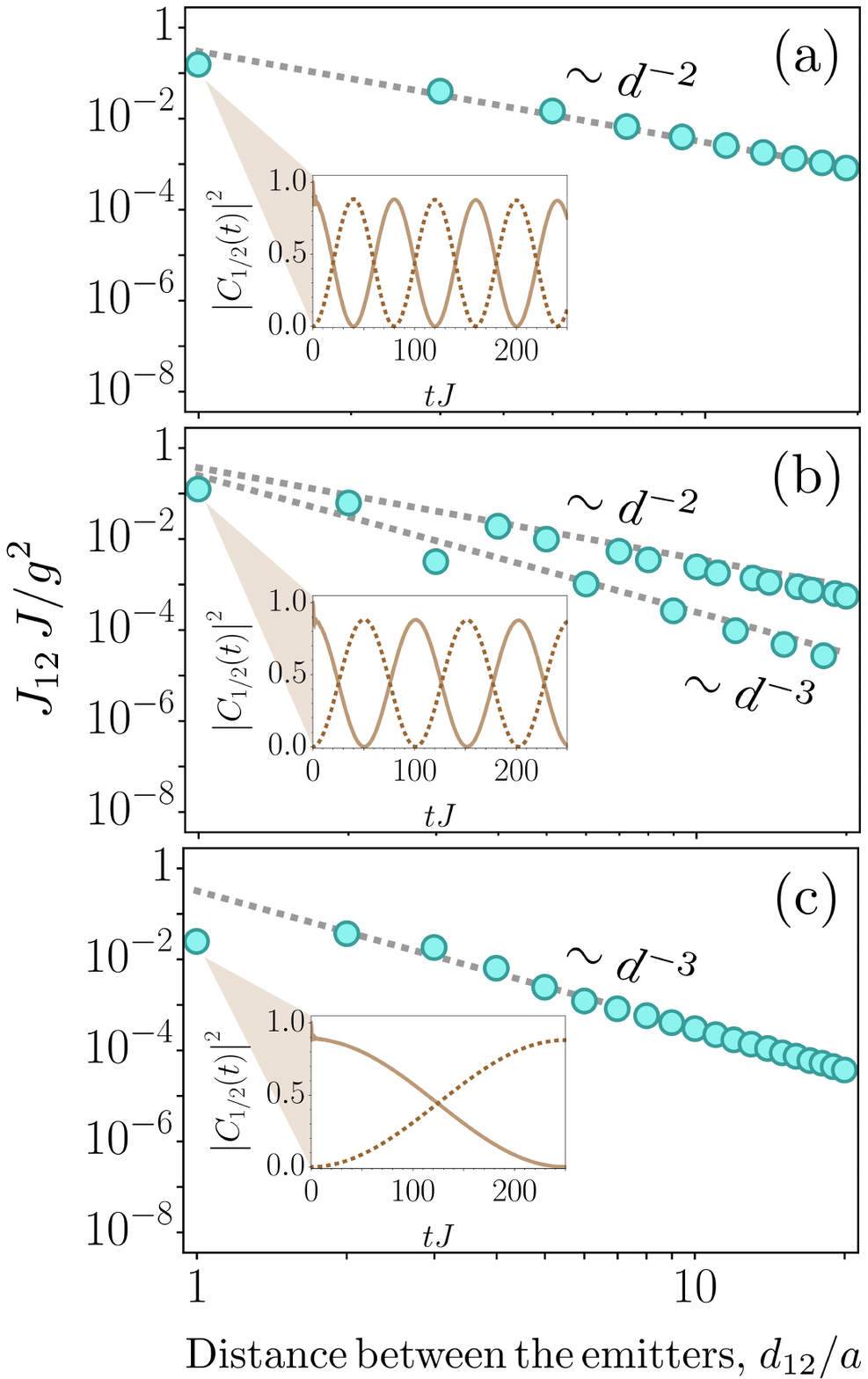}
     \vspace{0.0cm}
       \caption{
       (a), (b), (c): Frequency of the population's exchange experienced by two vertically aligned emitters as a function of the separation between them for $M=0,J$ and $2J$, respectively. Both emitters feature $\Delta=\Delta_c$ and are coupled to sites belonging to the $A$ sublattice. Gray dashed lines display the power law behavior associated to the Weyl bound state's confinement in the single emitter case. Insets show the dynamics of the excited state populations corresponding to two emitters ($|C_1(t)|^2$ and $|C_2(t)|^2$, solid and dotted lines, respectively), coupled to two different lattice sites with relative position $\rr_2-\rr_{1}=a\,\hat{z}$. In this calculation $g=0.5J$ is assumed.}
 \label{fig:Fig3}
 \vspace{-0.3cm}
 \end{figure}

The physical relevance of these tunable BS becomes most apparent when more than one emitter is coupled to the bath. In the following we focus on the case of two QEs ---initially prepared such that the first one is excited while the other is in its ground state-- and discuss exact calculations obtained through the application of the resolvent operator method to the case of two emitters~\cite{SIWeyl}. Insets of Figs.~\ref{fig:Fig3}(a)-(c) display the excited state population of the two emitters ($|C_{1,2}(t)|^2$) for $M=0$, $J$ and $2J$. The depicted dynamics correspond to a spatial configuration in which the two emitters, both of them coupled to a site belonging to the $A$ sublattice, are spaced a distance $a$ along the $z$ axis.  As seen, we obtain a set of coherent oscillations whose period increases with $M$. Remarkably, as shown in main panels of Figs.~\ref{fig:Fig3}(a)-(c), the dependence of the exchange frequency ($J_{12}$) with the vertical separation between emitters ($d_{12}$) replicate the power-law spatial confinement of the WBS photonic wavefunction found in the single emitter scenario. Analytical insight into these numerical results can be obtained by applying a perturbative (Markovian) approach to the configuration considered in Fig.~\ref{fig:Fig3}. Specifically, by formulating the problem in terms of a basis of symmetric and antisymmetric states, and making use of the vanishing density of states at the Weyl frequency, it can be shown~\cite{SIWeyl} that the temporal evolution of the excited state population of the emitters can be expressed in a simple closed form as $|C_{1/2}(t)|^2 \approx \left.\cos^2(J^M_{12}\:t)\:\Big/\: \sin^2(J^M_{12}\: t)\right.$, where $J^M_{12}$ is the excitation exchange frequency within the Markovian approximation. The difference between this analytical prediction and the numerical results of the non-perturbative approach shown in insets of Figs.~\ref{fig:Fig3}(a)-(c) lies in the fact that the oscillations predicted by the non-perturbative method are not complete, reaching a maximal value that coincides with the steady state population $|C_e(t\rightarrow \infty)|^2$ for each value of $M$. This, in turn, corroborates that the WBS is indeed mediating the interaction between the emitters.

\begin{figure}[t]
	\centering
	\includegraphics[width=7cm]{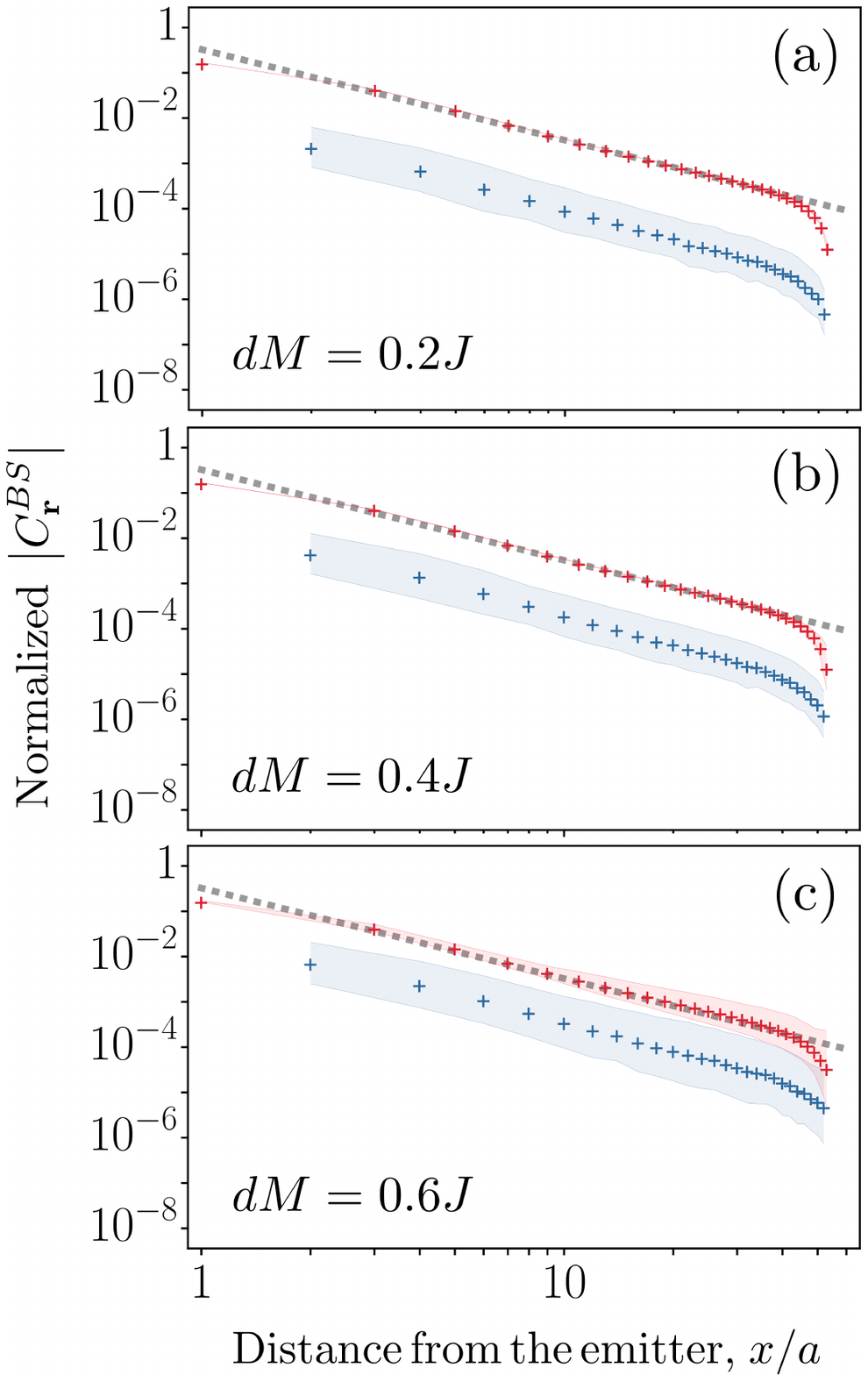}
	\caption{
		(a), (b), (c): Photonic component of the Weyl bound state (WBS) along the $x$-direction for $M=0$ in the presence of a random onsite energy disorder characterized by the strength value $dM=0.2J,0.4J$ and $dM=0.6J$, respectively. Blue/red crosses depict the average value of the normalized WBS's photonic component for A/B sites, while shadow areas span their corresponding standard deviation. For comparison, the gray dashed line displays the inverse square power law found in Fig.~3(a).
	}
	\label{fig:Fig4}
	\vspace{-0.3cm}
\end{figure}

Finally, we explore whether the above discussed topological protection of Weyl points translates into the WBS, and consequently to the interactions it mediates.~To do that, we study the robustness of the WBS against on-site energy disorder of the bath. Specifically, we consider a finite system with $\sim10^5$ sites and we focus on $M=0$ (a similar analysis can be performed for $M=J$ and $M=2J$). Then, we include diagonal energy disorder characterized by a strength value $dM$, so that the on-site energy of each localized mode is taken from a uniform random distribution within the range $[-dM,dM]$. Direct diagonalization of the system's Hamiltonian for $10^3$ random configurations and the subsequent statistical treatment yields the results shown in Fig.~4. For clarity, as the $M=0$ case is isotropic along the three Cartesian directions, we only display the $x$-direction. Blue/red crosses depict the average value of the normalized $|C_{\rr}^{WBS}|$ for A/B sites along the studied direction and shadow areas span their corresponding standard deviation. We observe that the original power-law behavior is maintained for the sites belonging to the $B$ sublattice with very small deviation up to $dM=0.6J$. In addition, we observe that, despite the fact that the projection of the WBS over sublattice $A$ sites vanishes in the non-disorder case, the presence of disorder introduces in those sites a similar power-law as that found for $B$ sites, accompanied by a significant increase of statistical deviation.

Summing up, we have unveiled the existence of a power-law, tunable, and robust to disorder atom-photon bound state appearing when quantum emitters interact with photonic Weyl environments with optical transitions close in energy to the Weyl points. Using an exact dynamical treatment, we have also evidenced that these Weyl bound states can mediate coherent interactions between the emitters with no associated dissipation, and that those interactions inherit the properties of the Weyl bound state. Beyond the particular realization reported here, we believe that the concept of using topological protected points to enable tunable, robust and long-range interactions could be exported to other setups to find different power-law behaviours, and, more generally, it could also stimulate further research of quantum optical phenomena in other topological photonic systems.

\acknowledgements{We acknowledge financial support from the Spanish MCIU under grant RTI2018-098452-B-I00 (MCIU/AEI/FEDER, UE) and the ``Mar\'ia de Maeztu'' programme for Units of Excellence in R\&D (MDM-2014-0377 and CEX2018-000805-M). AGT acknowledges support from CSIC Research Platform on Quantum Technologies PTI-001 and from Spanish project PGC2018-094792-B-100 (MCIU/AEI/FEDER, EU). IGE acknowledges financial support from the Ministry of Science and Innovation of Spain (FPU grant AP-2018-02748).}


\bibliography{books,Scigood}

\begin{thebibliography}{62}
\bibitem [{foo()}]{footnote44}%
  \BibitemOpen
  \href@noop {} {}\bibinfo {note} {Note that the studied Weyl photonic
  environment is actually classical, since it can be described by a quadratic
  Hamiltonian of bosonic modes. Only when it couples to the quantum emitters,
  which are non-linear objects, it can lead to purely quantum
  effects.}\BibitemShut {Stop}%
\end{thebibliography}

\end{document}